\input epsf.sty
\documentstyle[preprint,aps,floats]{revtex}
\draft
\begin{document}
\preprint{FNAL-PUB-95/153,hep-ex/9506007}
\title{
Limits on $\nu_\mu(\overline{\nu}_\mu)\to\nu_\tau(\overline{\nu}_\tau)$
and $\nu_\mu(\overline{\nu}_\mu)\to\nu_e(\overline{\nu}_e)$
Oscillations from a Precision Measurement of Neutrino-Nucleon
Neutral Current Interactions}
%\vglue 0.5cm
%\begin{sloppypar}
%\centering
%\noindent
\author{
        K.~S.~McFarland$^5$, D.~Naples$^5$,
        C.~G.~Arroyo$^4$, 
        P.~Auchincloss$^8$, P.~de~Barbaro$^8$
        A.~O.~Bazarko$^4$, R.~H.~Bernstein$^5$, A.~Bodek$^8$, T.~Bolton$^6$,
        H.~Budd$^8$, J.~Conrad$^4$, R.~B.~Drucker$^7$, 
        D.~A.~Harris$^8$, R.~A.~Johnson$^3$, J.~H.~Kim$^4$, B.~J.~King$^4$, 
        T.~Kinnel$^9$, G.~Koizumi$^5$, S.~Koutsoliotas$^4$,
        M.~J.~Lamm$^5$,W.~C.~Lefmann$^1$, 
        W.~Marsh$^5$,C.~McNulty$^4$, 
        S.~R.~Mishra$^4$, 
        P.~Nienaber$^{10}$, 
        M.~Nussbaum$^3$, 
        M.~J.~Oreglia$^2$, 
        L.~Perera$^3$, P.~Z.~Quintas$^4$,
        A.~Romosan$^4$,
        W.~K.~Sakumoto$^8$, 
        B.~A.~Schumm$^2$,
        F.~J.~Sciulli$^4$, W.~G.~Seligman$^4$, 
        M.~H.~Shaevitz$^4$, 
        W.~H.~Smith$^9$, P.~Spentzouris$^4$,
        R.~Steiner$^1$, E.~G.~Stern$^4$,
        M.~Vakili$^3$,U.~K.~Yang$^8$,
}
\address{
$^1$ Adelphi University, Garden City, NY 11530 \\
$^2$ University of Chicago, Chicago, IL 60637 \\
$^3$ University of Cincinnati, Cincinnati, OH 45221 \\
$^4$ Columbia University, New York, NY 10027 \\
$^5$ Fermi National Accelerator Laboratory, Batavia, IL 60510 \\
$^6$ Kansas State University, Manhattan, KS 66506 \\
$^7$ University of Oregon, Eugene, OR 97403 \\
$^8$ University of Rochester, Rochester, NY 14627 \\
$^9$ University of Wisconsin, Madison, WI 53706 \\
$^{10}$ Xavier University, Cincinnati, OH 45207 \\
}

\date{\today}

\maketitle

\newpage

\begin{abstract}
We present limits on 
$\nu_\mu(\overline{\nu}_\mu)\to\nu_\tau(\overline{\nu}_\tau)$ and
$\nu_\mu(\overline{\nu}_\mu)\to\nu_e(\overline{\nu}_e)$
oscillations based on a study of inclusive $\nu N$ interactions performed
using the CCFR massive coarse grained detector in the FNAL Tevatron Quadrupole
Triplet neutrino beam. The sensitivity to oscillations is
from the difference in the longitudinal energy deposition pattern
of $\nu_\mu N$ versus $\nu_\tau N$ or $\nu_e N$ charged current interactions.
The $\nu_\mu$ energies ranged from $30$ to $500$~GeV with a mean of $140$~GeV.  
The minimum and 
maximum $\nu_\mu$ flight lengths are $0.9$~km and $1.4$~km respectively.
For $\nu_\mu\to\nu_\tau$ oscillations, 
the lowest $90\%$ confidence upper limit in $\sin^22\alpha$ of 
$2.7\times 10^{-3}$ is obtained at $\Delta m^2\sim50$~eV$^2$.
This result is the most stringent limit to date for 
$25<\Delta m^2<90$~eV$^2$.
For $\nu_\mu\to\nu_e$ oscillations, 
the lowest $90\%$ confidence upper limit in $\sin^22\alpha$ of 
$1.9\times 10^{-3}$ is obtained at $\Delta m^2\sim350$~eV$^2$.
This result is the most stringent limit to date for 
$250<\Delta m^2<450$~eV$^2$, and also excludes at $90\%$ confidence
much of the high $\Delta m^2$ region favored by the recent LSND
observation.
\end{abstract}

 The mixing of non-degenerate neutrino mass eigenstates 
would lead to oscillations of one neutrino type into another.
For mixing between two generations, the oscillation probability is 
\begin{equation}
P(\nu_1\to\nu_2)=\sin^22\alpha \sin^2 \left ( \frac{\Delta m^2L}{E_\nu}
\times 1.27{\rm\textstyle\frac{GeV}{km\;eV^2}}\right ) ,
\label{osc}
\end{equation}
where $\Delta m^2$ is $|m_1^2-m_2^2|$, $\alpha$ is 
the mixing angle, $E_\nu$ is the neutrino energy, and L is the distance
the neutrino travels between production and observation.
Nonzero neutrino mass and mixing would have
important implications for cosmology and particle physics.
Neutrino oscillations may explain 
observed neutrino deficits from the sun and from atmospheric sources.
%Accelerator experiments will continue the 
%search for neutrino mixing with particular emphasis on the tau sector
%\cite{tau} \cite{minos}.

To date the best limits for $\nu_\mu\to\nu_\tau$ oscillations are
derived from searches for $\nu_\tau$ appearance through exclusive
processes.  For example, the FNAL-E531 limit \cite{E531} comes from
searching for a detached vertex from a tau decay in emulsion.  A fine
grained detector ({\it i.e.} emulsion, or a low density fine grained
calorimeter such as was used by CHARM II \cite{CHARMII}) is necessary
to be sensitive to low mixing angles through exclusive modes.  In the
case of $\nu_\mu\to\nu_e$ oscillations, the best limits from
accelerator experiments come from fine-grained calorimetric (e.g.:
BNL-E734\cite{BNL734}, BNL-E776\cite{BNL776}) or fully active
detectors (e.g.: KARMEN\cite{KARMEN}, LSND\cite{LSND}, searching for
quasi-elastic charged-current production of electrons.  The LSND
experiment, using a liquid scintillator neutrino target, has recently
reported a signal consistent with
$\overline{\nu_\mu}\to\overline{\nu_e}$ at a $\sin^22\alpha$ of
$\sim10^{-2}$ and a $\Delta m^2\stackrel{>}{\sim} 1$~eV${^2}$
\cite{LSND}.

In this report, we obtain results with comparable sensitivity to E531,
CHARM II for $\nu_\mu\to\nu_\tau$ oscillations and to BNL-E734,
BNL-E766 and KARMEN for $\nu_\mu\to\nu_e$ oscillations at $\Delta
m^2\stackrel{>}{\sim}40$~eV$^2$ using the massive and relatively
coarse grained CCFR detector.  The main advantage of this type of
detector is increased interaction probability which will be 
particularly important in
a low flux, long baseline neutrino beam \cite{minos}.  Our result
establishes the sensitivity of such detectors to small mixing angles.

The CCFR detector \cite{NIMpaper} consists of an $18$~m long,
$690$~ton target calorimeter with a mean density of $4.2$~g/cm$^3$,
followed by an iron toroid spectrometer.  The target calorimeter
consists of 168 iron plates, $3$m~$\times$~$3$m~$\times$~$5.1$cm each. 
The active elements are liquid scintillation counters spaced
every two plates and drift chambers spaced every four plates.
There are a total of 84 scintillation counters and 42 drift chambers
in the target. The toroid spectrometer is not directly used in this
analysis.

\begin{figure}
\epsfxsize=\textwidth
\epsfbox{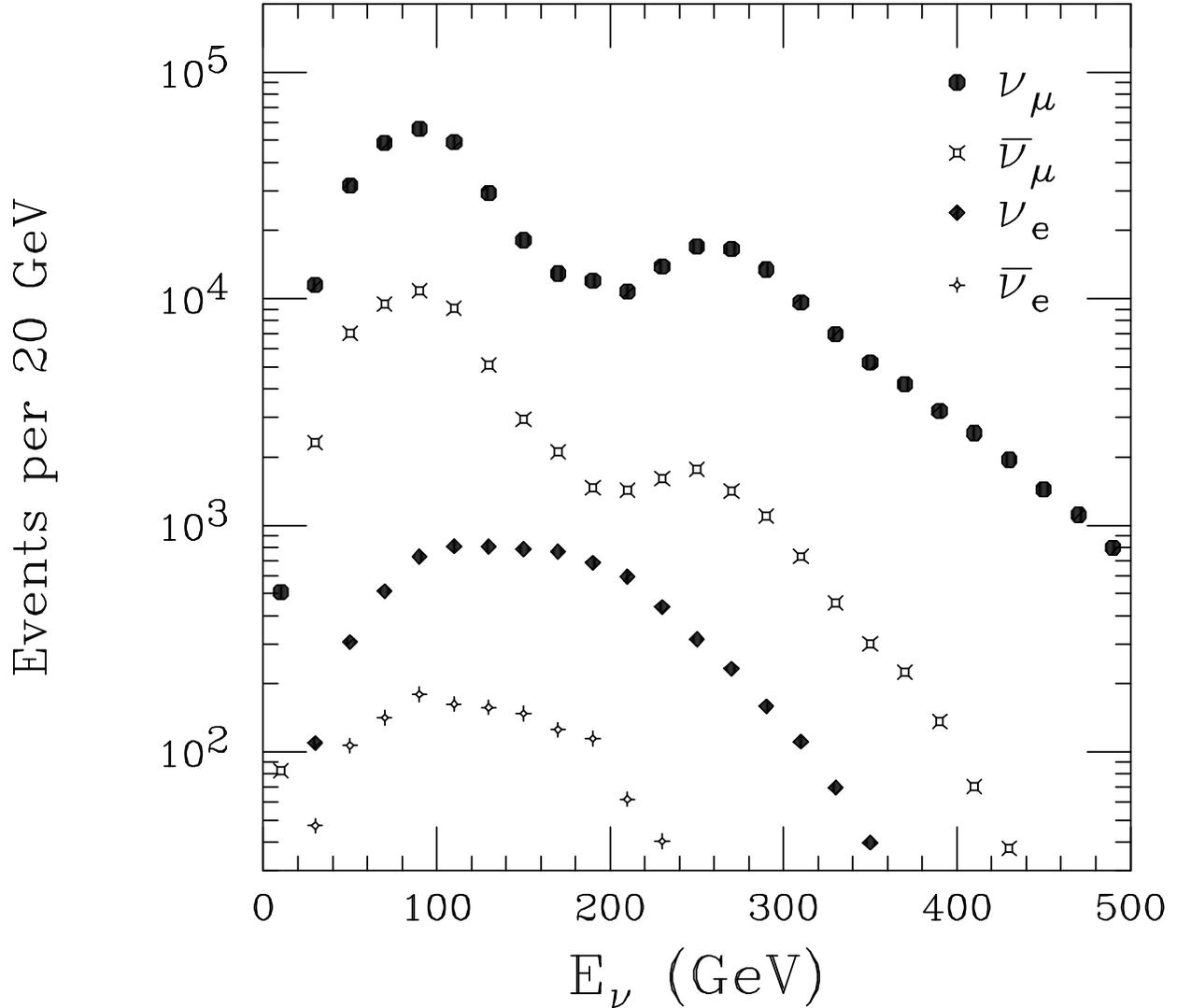}
\caption{
Neutrino energy spectra for $\nu_\mu$, 
$\overline{\nu}_\mu$, $\nu_e$, and $\overline{\nu}_e$ 
at the CCFR detector for the FNAL wideband 
neutrino beam. (Monte Carlo based on measured 
relative $\nu_\mu$ and $\overline{\nu}_\mu$ fluxes).}
\label{fig:enu}
\end{figure}

The Tevatron Quadrupole Triplet neutrino beam is created by decays of
pions and kaons produced when $800$~GeV protons hit a production target.
A wide band of secondary energies is accepted by focusing magnets. 
The production target is located about $1.4$~km upstream of the
neutrino detector. The production target and focusing train are
followed by a $0.5$~km decay region. The resulting
neutrino energy spectra for $\nu_\mu$, $\overline{\nu}_\mu$, $\nu_e$,
and $\overline{\nu}_e$ at the detector are shown in Figure
\ref{fig:enu}.
The beam contains a $2.3$\% fraction of electron neutrinos
and a negligible fraction of tau neutrinos (less than $10^{-5}$)
which result primarily from $D_s$ decay.

Neutrinos are observed in the target calorimeter {\it via} their neutral
current and charged current interactions.  $\nu_\mu$ charged current events
are characterized by the presence of a muon in the final
state which deposits energy in a large number of consecutive scintillation
counters as it travels through the calorimeter.  Neutral current
events have no muon and deposit energy over a range of counters
typical of a hadronic shower (5 to 20 counters).  Accordingly,
we define ``short'' events as those which deposit energy over an
interval of 30 or fewer scintillation counters.
The ratio $R_{30}$ is defined to be the number of
short events divided by the number of long events \cite{WMA}.  
This ratio is strongly dependent on the ratio of neutral to charged current
events which is a function of the electroweak mixing angle, $\sin^2\theta_W$.

Assuming the validity of the Standard Model, $\sin^2\theta_W$ is
accurately measured from other processes.  We can use these
measurements to predict the ratio of neutral to charged current events
in the CCFR detector, and thus $R_{30}$.  The presence of $\nu_\tau$
or additional $\nu_e$
in the neutrino beam would cause the measured $R_{30}$ to be larger
than its calculated value because most charged current tau and
electron neutrino
interactions do not produce a muon in the final state and will thus
appear ``short''.

In this study, we attribute any deviation in our measured $R_{30}$
from the predicted value to $\nu_\mu\to\nu_\tau$  or
$\nu_\mu\to\nu_e$ oscillations.  This technique, which has
been discussed previously in the literature 
\cite{minos}\cite{Loverre}\cite{ref:bob}, assumes that only
one of the two types of flavor oscillation contributions to a change
in $R_{30}$, and is therefore conservative since both types of
oscillation would increase the measured $R_{30}$.

We used a detailed Monte Carlo to relate a given $\nu_\mu\to\nu_\tau$
or $\nu_\mu\to\nu_e$
oscillation probability to the quantity $R_{30}$.  The 
$\sin^2\theta_W$ value\footnote {This value for $\sin^2\theta_W$
is obtained using the world average value $M_W$ measurement
\cite{ref:mw}, the prediction from the measured $M_Z$, and the average of 
all LEP and SLD Z-pole measurements from \cite{ref:langacker}.  The
$M_Z$ extraction is corrected for the recent re-evaluation of
$\alpha_{EM}(M_Z^2)$ by Swartz\cite{ref:swartz}.  A top mass of 
$180\pm12$~GeV\cite{topmass} and $60<M_{\rm Higgs}<1000$~GeV are used 
to convert from the $\overline{\rm\textstyle MS}$ and $M_Z$ schemes 
to the on-shell scheme used here.} in the on-shell renormalization 
scheme of $0.2232\pm 0.0018$ is input to the Monte Carlo.
The other inputs to the Monte Carlo are 
parameterizations of the measured CCFR detector responses \cite{NIMpaper}, 
nucleon structure functions \cite{sf}, and relative neutrino beam fluxes 
extracted from the charged current data sample \cite{flux}. 
The $\nu_e$ flux is modeled in a detailed beamline simulation,
normalized by the observed $\nu_\mu$ flux \cite{WMA}.
The same beamline simulation is used to tag the decay location for each 
pion and kaon and thus the creation point of each $\nu_\mu$ 
along the beamline. 
The measured flux gives the number of $\nu_{\mu}$'s at the 
detector. $P(\nu_\mu\to\nu_{\tau,e})$ is determined from
Eq.~\ref{osc} and the beamline simulation.
We assume 
$P(\nu_\mu\to\nu_{\tau,e})=P(\overline{\nu}_\mu\to\overline{\nu}_{\tau,e})$
(a consequence of CP invariance).
The number of $\nu_\mu$'s produced in the beamline is then
the number observed at the detector divided by 
$1-P(\nu_\mu\to\nu_{\tau,e})$.
The predicted electron neutrino flux is rescaled to the 
{\em produced} number of 
$\nu_\mu$'s.  The tau or electron neutrino flux from
neutrino oscillations is calculated by multiplying the
{\em produced} number of $\nu_\mu$'s by $P(\nu_\mu\to\nu_{\tau,e})$.

To simulate $\nu_{\tau,e}$ interactions in our detector we assumed the
$\nu_{\tau,e}$ neutral current cross section is the same as for $\nu_\mu$
interactions. The $\nu_\tau$ charged current cross section was
calculated including mass suppression terms.  Following \cite{heavy}
we used the approximation that the structure functions $F_4=0$, and
$xF_5=2xF_1$.
The kinematic suppression for massive particle production was also
taken into account. The Monte Carlo program TAUOLA \cite{tauola} was
used to simulate tau decays.
We define $E_{cal}$ as the energy deposited in the calorimeter in the
first twenty counters following the event vertex.  For $\nu_{\tau,e}$
charged current events $E_{cal}$ includes the visible energy from the
tau decay.  
Events are required to deposit a minimum energy of 30 GeV in the target 
calorimeter. The contributions from quasi-elastic and resonance 
production are suppressed by this requirement.

\begin{figure}
\epsfxsize=\textwidth 
\epsfbox{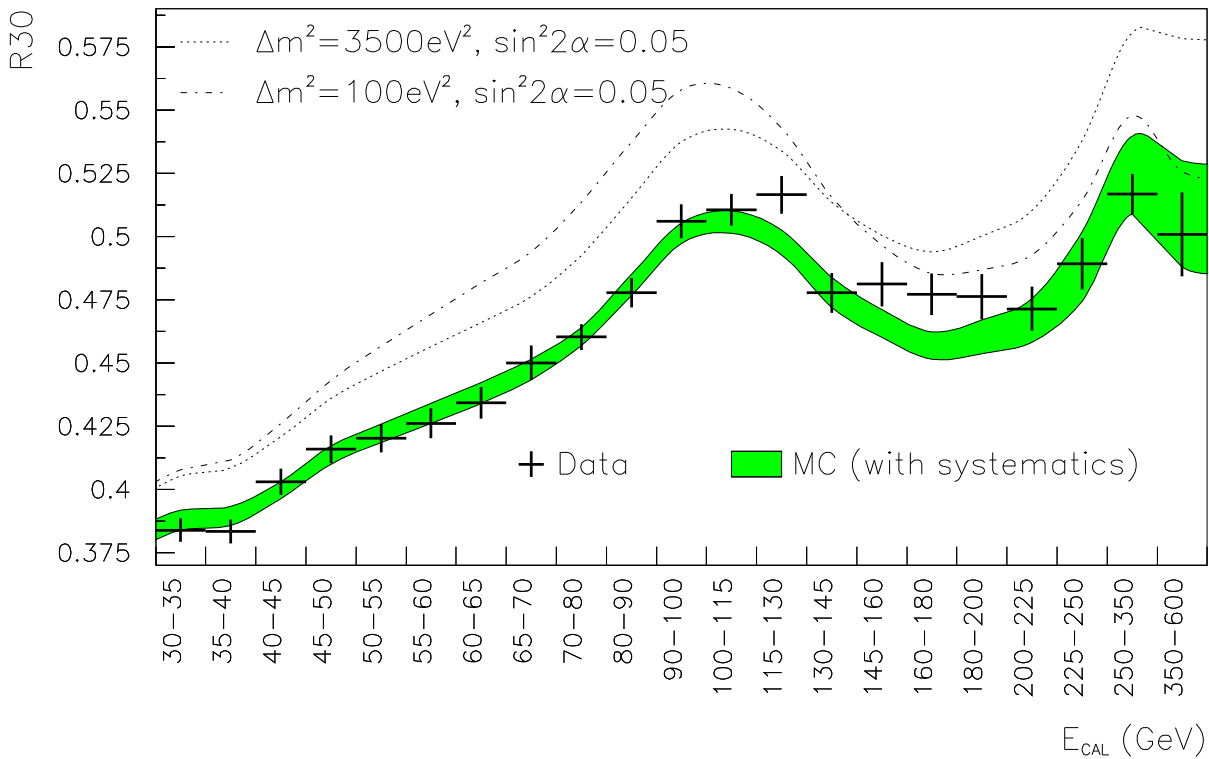}
\caption{
$R_{30}$ as a function of $E_{cal}$ for the data (points). The filled
band shows Monte Carlo assuming no oscillations
with 1$\sigma$ systematic errors added in quadrature. 
Data points show statistical errors only.
The dotted curve corresponds to $\nu_\mu\to\nu_\tau$ oscillations
with $\Delta m^2=3500$ eV$^2$ and 
$\sin^22\alpha=0.05$ and the dashed curve to
$\Delta m^2=100$ eV$^2$ and $\sin^22\alpha=0.05$.
The curve for $\Delta m^2=100$ eV$^2$ corresponds to lower energy neutrinos 
for which less energy is deposited in the calorimeter. At high
$\Delta m^2$ the high $E_{cal}$ events are most sensitive to 
$\nu_\tau$ or $\nu_e$ appearance.}
\label{fig:r30}
\end{figure}

Events were selected using a calorimeter trigger fully sensitive for
$E_{cal}$ above 20 GeV.  To ensure event containment,
the fiducial volume of the detector is limited to
a central cylindrical region 30'' in radius and excludes events which began in
the first 6 counters or the last 34 counters of the detector.
The resulting data sample consisted of about 450,000 events. 
The data and Monte Carlo are divided into 21 $E_{cal}$ bins. 
For each $\Delta m^2$, 
the Monte Carlo prediction for $R_{30}(E_{cal},\sin^22\alpha)$
is compared with $R_{30}(E_{cal})$ from the data.
Figure \ref{fig:r30} shows the $R_{30}$ distribution as a
function of $E_{cal}$ for the data and for the Monte Carlo
simulation. The detailed shape of $R_{30}(E_{cal})$ 
depends on many competing effects which are put into
the Monte Carlo, but is dominated by the 
variation of short charged current events 
with $E_{cal}$ and by the contribution from the predicted $\nu_e$ flux.

\begin{table}
%\squeezetable
  \begin{tabular}{|l|l|l|l|} 
Source of Error &  $\Delta m^2=3500$~eV$^2$ & $310$~eV$^2$ & $80$~eV$^2$ \\ \hline 
statistical& $2.4\times10^{-3}$& $1.8\times10^{-3}$& $2.1\times10^{-3}$  \\ 
$\nu_e$ beam content & $2.5\times10^{-3}$& $2.0\times10^{-3}$& $1.4\times10^{-3}$ \\
detector systematics & $2.2\times10^{-3}$ & $2.0\times10^{-3}$ & $2.0\times10^{-3}$ \\
charm mass &  $2.4\times10^{-3}$& $1.6\times10^{-3}$& $2.4\times10^{-3}$ \\
charm sea &  $1.2\times10^{-3}$& $0.8\times10^{-3}$& $1.0\times10^{-3}$ \\
$\sin^2\theta_W$ & $1.7\times10^{-3}$ & $1.2\times10^{-3}$ & $1.5\times10^{-3}$ \\
other model & $0.6\times10^{-3}$ & $0.6\times10^{-3}$ & $0.7\times10^{-3}$ \\
Total& $5.2\times10^{-3}$ & $4.1\times10^{-3}$ & $4.4\times10^{-3}$ \\
  \end{tabular}
  \caption{The change in $\sin^22\alpha$ for $\nu_\mu\to\nu_\tau$ 
from a one sigma shift in
the dominant errors. The row labeled ``total'' 
includes these and many smaller uncertainties added in quadrature.}
  \label{errors}
\end{table}

There are four major uncertainties in the comparison of $R_{30}(E_{\rm cal})$
from the Monte Carlo to the data: the statistical error in
the data, the uncertainty in the effective charm quark mass for
charged current charm production, the uncertainty in the incident flux of
$\nu_e$'s on the detector, and the uncertainty in the on-shell weak
mixing angle from outside measurements.  Other sources of systematic
uncertainty  were also investigated
\cite{WMA}.   Table~\ref{errors} shows the effect of
the uncertainties for three choices of $\Delta m^2$.

The charm mass error comes from the uncertainty in modeling the turn-on
of the charm quark production cross section.  The Monte
Carlo uses a slow-rescaling model with the parameters extracted
using events with two oppositely charged muons in this experiment \cite{baz}. 
This error dominates the calculation of $R_{30}$ at low $E_\nu$ (and 
low $E_{cal}$) where the threshold suppression is
greatest.  The $\nu_e$ flux uncertainty has a large effect on $R_{30}$
because almost all charged current $\nu_e$ events are short events.
Therefore, the relatively small ($4.2\%$ \cite{WMA}) 
fractional uncertainty in the $\nu_e$
flux is a large effect, particularly at high $E_{cal}$ since most
$\nu_e$ charged current interactions deposit the full incident
neutrino energy into the calorimeter.  
This $4.2\%$ is 
dominated by a $20\%$ production uncertainty in the $K_L$ content of
the secondary beam which produces $16\%$ of the $\nu_e$ flux.  The bulk of the
$\nu_e$ flux comes from $K^{\pm}_{e3}$ decays, which are well-constrained
by the observed $\nu_\mu$ spectrum from $K^{\pm}_{\mu 2}$ decays \cite{WMA}.

%The data are fit by forming a $\chi^2$ function given by:
%\begin{eqnarray*}
%\lefteqn{\chi^2=\sum_{\rm systematics}C_i^2 }\\
%&&+\sum_{\rm E_{\rm cal}}\left( \frac{{\rm Data}(E_{\rm cal})-
%                 {\rm MC}(E_{\rm cal}; \Delta m^2, \sin^22\alpha;
%C_i)}{\sigma_{\rm stat}}\right ) ^2  .
%\end{eqnarray*}
%This incorporates the Monte Carlo generated effect of the oscillations
%as well the effect of the systematic errors.
%The systematic coefficients, $C_i$, represent 
%deviations of systematic effects
%from their nominal value in units of systematic sigmas.
The data are fit by forming a $\chi^2$ which incorporates
the Monte Carlo generated effect of oscillations, and statistical and
systematic uncertainties.
A best fit $\sin^22\alpha$ is determined for each 
$\Delta m^2$ by minimizing the $\chi^2$ as a function of 
$\sin^22\alpha$ and the 33 systematic coefficients, $C_i$.
Best fit values of $\sin^22\alpha$ with one sigma errors from 
the fit are shown in Tables~\ref{frequentist} and \ref{frequentistE}.
At all $\Delta m^2$, the data are consistent with no observed
$\nu_\mu\to\nu_{\tau,e}$ oscillation. The statistical significance
of the best-fit oscillation at any $\Delta m^2$ is at most 1.2 sigma. 

\begin{table}
%\squeezetable
  \begin{tabular}{|c|c|c||c|c|c|}
$\Delta m^2$ (eV$^2$) & Best Fit & Sigma & 
$\Delta m^2$ (eV$^2$) & Best Fit & Sigma \\
\hline 
    2.0&-2.1114&2.0192&185.0& 0.0050&0.0042\\
    3.5&-0.6982&0.6676&  200.0& 0.0047&0.0042\\
    5.0&-0.3419&0.3268&  220.0& 0.0040&0.0041\\
    6.0&-0.2373&0.2267&   240.0& 0.0033&0.0041\\
    8.0&-0.1351&0.1296&  275.0& 0.0022&0.0040\\
   10.0&-0.0872&0.0838&  295.0& 0.0018&0.0040\\
   15.0&-0.0397&0.0385&  310.0& 0.0016&0.0040\\
   20.0&-0.0229&0.0224&  350.0& 0.0010&0.0041\\
   35.0&-0.0084&0.0090&  400.0& 0.0002&0.0043\\
   42.0&-0.0061&0.0069&  430.0&-0.0002&0.0044\\
   50.0&-0.0045&0.0056&  500.0&-0.0002&0.0047\\
   60.0&-0.0031&0.0047&  550.0& 0.0004&0.0048\\
   70.0&-0.0021&0.0041&  600.0& 0.0013&0.0050\\
   80.0&-0.0013&0.0038&  650.0& 0.0020&0.0051\\
   90.0&-0.0006&0.0036&  700.0& 0.0026&0.0051\\
  100.0& 0.0001&0.0035&  750.0& 0.0028&0.0050\\
  110.0& 0.0008&0.0035&  800.0& 0.0027&0.0049\\
  120.0& 0.0015&0.0036& 1000.0& 0.0017&0.0049\\
  135.0& 0.0027&0.0037& 2000.0& 0.0018&0.0050\\
  150.0& 0.0038&0.0039& 3500.0& 0.0018&0.0049\\
  175.0& 0.0049&0.0041&10000.0& 0.0018&0.0049\\
  \end{tabular}
  \caption{The result for $\sin^22\alpha$ from the fit at
each $\Delta m^2$ for $\nu_\mu\to\nu_\tau$ oscillations}
  \label{frequentist}
\end{table}
\begin{table}
%\squeezetable
  \begin{tabular}{|c|c|c||c|c|c|}
$\Delta m^2$ (eV$^2$) & Best Fit & Sigma & 
$\Delta m^2$ (eV$^2$) & Best Fit & Sigma \\
\hline 
    2.0&-0.7856& 1.0714&  200.0& 0.0008& 0.0021\\
    3.0&-0.3552& 0.4825&  225.0& 0.0001& 0.0020\\
    4.0&-0.1993& 0.2720&  250.0&-0.0001& 0.0019\\
    5.0&-0.1275& 0.1746&  275.0&-0.0001& 0.0017\\
    7.0&-0.0650& 0.0897&  300.0&-0.0001& 0.0016\\
    9.0&-0.0392& 0.0548&  350.0&-0.0001& 0.0015\\
   10.0&-0.0317& 0.0448&  400.0& 0.0000& 0.0016\\
   20.0&-0.0074& 0.0122&  450.0& 0.0004& 0.0018\\
   30.0&-0.0027& 0.0061&  500.0& 0.0009& 0.0020\\
   40.0&-0.0009& 0.0039&  600.0& 0.0015& 0.0024\\
   50.0&-0.0001& 0.0028&  700.0& 0.0014& 0.0025\\
   60.0& 0.0004& 0.0023&  800.0& 0.0007& 0.0023\\
   70.0& 0.0007& 0.0019& 1000.0& 0.0009& 0.0023\\
   80.0& 0.0009& 0.0018& 1500.0& 0.0009& 0.0022\\
   90.0& 0.0012& 0.0017& 2000.0& 0.0009& 0.0023\\
  100.0& 0.0014& 0.0017& 5000.0& 0.0009& 0.0023\\
  125.0& 0.0022& 0.0018&10000.0& 0.0009& 0.0023\\
  150.0& 0.0024& 0.0020&20000.0& 0.0009& 0.0023\\
  175.0& 0.0017& 0.0021&50000.0& 0.0009& 0.0023\\
  \end{tabular}
  \caption{The result for $\sin^22\alpha$ from the fit at
each $\Delta m^2$ for $\nu_\mu\to\nu_e$ oscillations}
  \label{frequentistE}
\end{table}
\begin{figure}
\epsfxsize=\textwidth 
\epsfbox{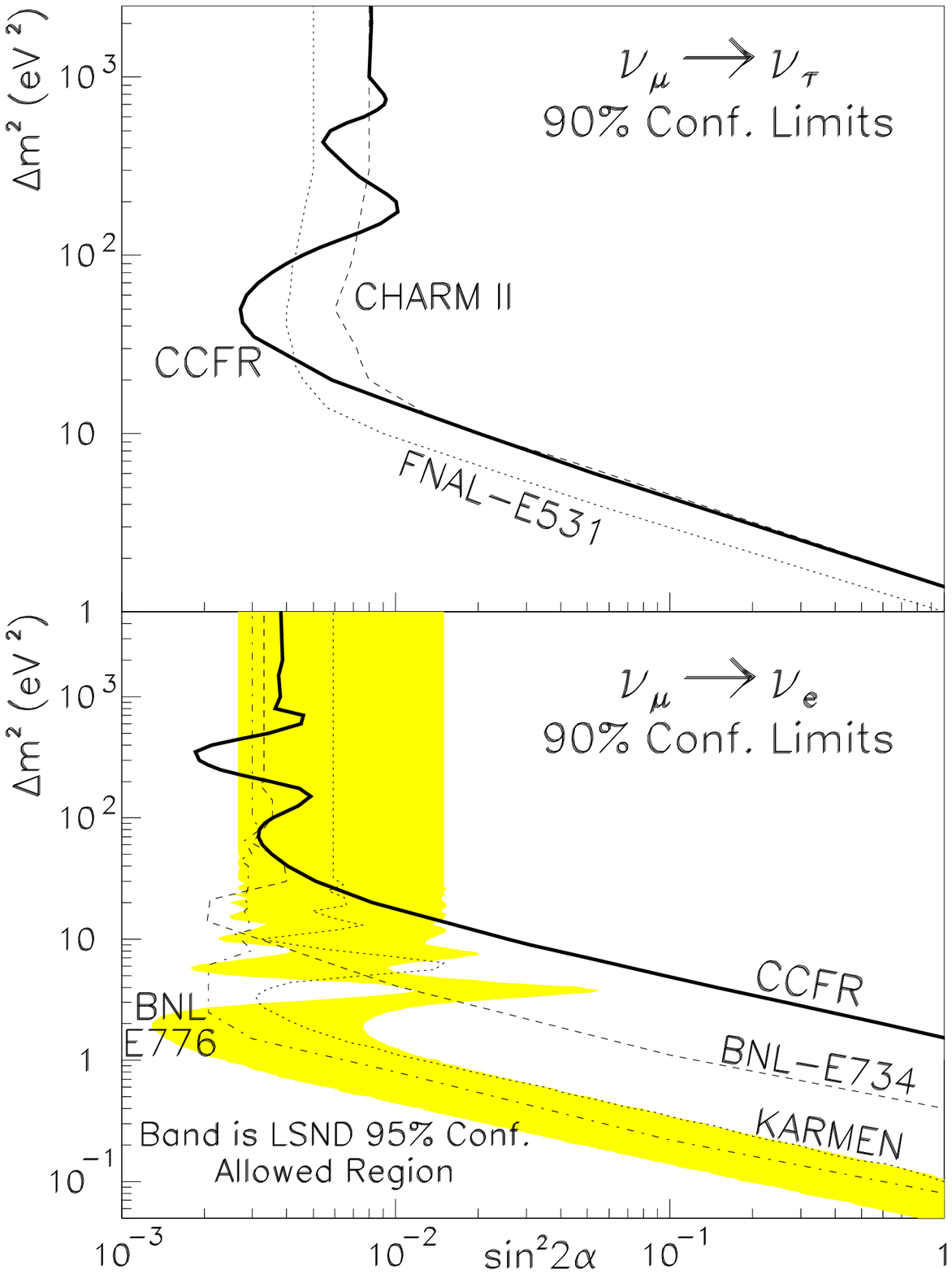}
\caption{Excluded region of $\sin^2 2\alpha$ and $\Delta m^2$
for $\nu_\mu\to\nu_\tau,e$ oscillations 
from this analysis at $90\%$ confidence is shown as
dark, solid curves.}
\label{fig:result}
\end{figure}

The frequentist approach
\cite{pdg:freq} is used to set a 90\% confidence upper limit for each 
$\Delta m^2$.  The limit in $\sin^22\alpha$ at each $\Delta m^2$
corresponds to a shift of $1.64$ units of $\chi^2(\sin^22\alpha)$ from
the minimum $\chi^2$ (at the best fit value in
Table~\ref{frequentist}).  The $\chi^2$ value for the no-oscillations
case is 15.7/21 dof.  The 90\% confidence upper limit is plotted in
Figure~\ref{fig:result} for $\nu_\mu\to\nu_\tau$.  The best limit of
$\sin^22\alpha<2.7\times10^{-3}$ is at $\Delta m^2 = 50$~eV$^2$.  For
$\sin^22\alpha=1$, $\Delta m^2 > 1.4$~eV$^2$ is excluded, and for
$\Delta m^2\gg1000$~eV$^2$, $\sin^22\alpha>8.1\times10^{-3}$ is
excluded at $90\%$ confidence.  For $\nu_\mu\to\nu_e$ oscillations,
the 90\% confidence upper limit is also shown in Figure~\ref{fig:result}.
The best limit of $\sin^22\alpha<1.9\times10^{-3}$ is at $\Delta m^2 =
350$~eV$^2$.  For $\sin^22\alpha=1$, $\Delta m^2 > 1.6$~eV$^2$ is
excluded, and for $\Delta m^2\gg1000$~eV$^2$,
$\sin^22\alpha>3.8\times10^{-3}$

This result demonstrates sensitivity to low mixing angles in a high mass,
coarse grained sampling calorimeter and has implications for
proposed long-baseline experiments \cite{minos}.  However, a
detailed Monte Carlo study of the sensitivity of those experiments
must be performed to correctly apply this result.  The lower
energy and the lower level of statistics in the long baseline
experiments will result in less statistical sensitivity, while having
both a near and far detector would reduce many of the other sources of
uncertainty \cite{ref:bob} listed in Table~\ref{errors}.

In conclusion, we have used a new analysis method to search for
$\nu_\mu\to\nu_{\tau,e}$ oscillations with a coarse-grained
calorimetric detector.  We see a result consistent with no
neutrino oscillations and find $90\%$ confidence level excluded
regions in $\sin^2 2\alpha$-$\Delta m^2$ space. This result is the most 
stringent limit to date for $\nu_\mu\to\nu_\tau$ oscillations with 
$25<\Delta m^2<90$~eV$^2$ and for $\nu_\mu\to\nu_e$ oscillations with 
$250<\Delta m^2<400$~eV$^2$.

\end{document}